\documentclass[twocolumn,showpacs,preprintnumbers,amsmath,amssymb]{revtex4}

\usepackage{xcolor}
\usepackage{graphicx}
\usepackage{graphicx,epstopdf}
\usepackage{dcolumn}
\usepackage{bm}
\usepackage{mathrsfs}

\begin{document}


\title{Emergence of synchronization in multiplex networks of mobile R\"{o}ssler oscillators}

\author{Soumen Majhi$^1$}
\author{Dibakar Ghosh$^1$}
\email{dibakar@isical.ac.in}
\author{J\"{u}rgen Kurths$^{2,3}$}
\affiliation{$^1$Physics and Applied Mathematics Unit, Indian Statistical Institute, 203 B. T. Road, Kolkata-700108, India\\
$^2$ Potsdam Institute for Climate Impact Research, Potsdam 14473, Germany\\
$^3$ Saratov State University, Saratov, Russia}
\date{\today}

\begin{abstract}
Different aspects of synchronization emerging in dynamical networks of coupled oscillators have been examined prominently in the last decades. Nevertheless, little attention has been paid on the emergence of this imperative collective phenomenon in networks displaying temporal changes in the connectivity patterns. However there are numerous practical examples where interactions are present only at certain points of time owing to physical proximity. In this work, we concentrate on exploring the emergence of interlayer and intralayer synchronization states in a multiplex dynamical network comprising of layers having mobile nodes performing a random walk. We thoroughly illustrate the impacts of the network parameters, in particular the vision range $\phi$ and the step size $u$ together with the inter- and intralayer coupling strengths $\epsilon$ and $k$ on these synchronous states arising in coupled R\"{o}ssler systems. The presented numerical results are very well validated by analytically derived necessary conditions for the emergence and stability of the synchronous states. Furthermore, the robustness of the states of synchrony is studied under both structural and dynamical perturbations. We find interesting results on interlayer synchronization for a continuous removal of the interlayer links as well as for progressively created static nodes. We demonstrate that the mobility parameters responsible for intralayer movement of the nodes, can retrieve interlayer synchrony under such structural perturbations. For further analysis of survivability of interlayer synchrony against dynamical perturbations, we proceed through the investigation of single-node basin stability, where again the intralayer mobility properties have noticeable impacts. We also discuss the scenarios related mainly to effects of the mobility parameters in terms of percolation of the whole network.

\end{abstract}


\maketitle


\section{Introduction}
Complex network theory has offered a valuable platform for the study of emergent collective behaviors, particularly based on the coaction in the interactional topology and coupled dynamical units. Among the diverse emerging asymptotic states, the developing phenomenon of {\it synchronization} \cite{syn_kurths,sycha} has been in the center of research since the last two decades. Synchronization refers to a mechanism in which interacting systems adjust their rhythm over time to achieve a coherent behavior. Synchronization processes are omnipresent in nature and play a crucial role in many different contexts as biology, ecology, climatology, sociology, technology etc. \cite{syn_kurths,sycha,syn_stro}.
\par But notably, most of the previous works dealt with situations where the links between the local dynamical units are presumed to be persistent for all the course of time. However, this sort of simplified assumption may not resemble many practical situations. Indeed, there exists various examples of networks in real world encircling time evolving interaction structure, where the links associating the nodes vary over time. Networks with time-varying topology  \cite{tempholm} may arise in physical systems \cite{pts}, biological systems \cite{fbn}, communication networks \cite{mcom,mcom2,mcom3,mcom4}, social networks \cite{cons}, modeling of spread of epidemics \cite{infectt,infect,sinha3}, to name only a few. This implies that a shift from the static assumption to dynamical evolution of a network itself is essential for further understanding. In recent times, the study of synchronization in time-varying networks has secured growing attention, particularly, in networks where variation in interaction arises due to mobility in the nodes. In fact, studies have revealed that there are enormous practical evidences in which order emerges among a large number of movable individuals that can interact with each other strictly depending on their proximity. Coordination arising in an ensemble of mobile robots or vehicles was proposed as a procedure of controlling them \cite{robot,vehicle}. A transition from disorder to spontaneous order in moving swarming desert locusts was explored by Buhl et al. \cite{grani}. Authors of \cite{cell1} discussed the issue of cell movement involved in local intercellular signaling which is highly important for morphogenesis during animal development. Particularly, influence of collective cell movement in enhancing synchronization of locally coupled genetic oscillators, explaining synchrony of the segmentation clock in zebrafish somitogenesis was also illustrated \cite{cell2}. Synchrony among mobile agents is also crucial in several other developments, such as the process of chemotaxis \cite{chemo}, mechanisms in wireless sensor networks, MANET (Mobile Ad Hoc Network) etc. \cite{sensor,adhoc}. \par These bring the essentiality of the study of synchronization in dynamical networks of mobile nodes. Regarding this, there exists few works in the literature, e.g., the study in \cite{infect} particularly focussed on the spread of infectious disease while considering a network of movable oscillators. Ensemble of mobile Kuramoto phase oscillators and chaotic systems are respectively dealt with in \cite{movkur,movkur2} and \cite{movch,chaosm}, in order to analyze the emergence of synchrony. Synchrony in a dynamical network of nodes each associated with integrate-and-fire oscillators communicating with immediate neighbors upon firing time, was elaborately discussed in \cite{if1}. The impacts of noise \cite{mornoise,chaos2016}, a typical restricted interaction \cite{kim} on the collective global synchronous behavior have also been investigated. Other notable works include studies about the effect of self-propulsion on synchrony in mobile oscillators' network \cite{rev2,rev1}.
\par We would now like to emphasize that recent research also corroborated that the functionalities arising in one network may awfully influence other networks and particularly, a node in one network is quite likely a part of another network as well. From social \cite{multi_soc} and biological systems \cite{neuronal} to physical and transportation systems \cite{trans}, it is clear that such interrelationships exist from different contexts. This proves multilayer (multiplex) architecture of networks \cite{kivela,boccal,gomez,mathemat,nicosia,buldyrev,gao,helbing,podob,podobnik} to be quite effective in describing many systems. This also suggests that the study of dynamical evolution processes on top of multilayer networks, which massively differ from those on its monolayer counter parts \cite{cardillo}, are the next frontier in network theory research. In this context, one must also note that there are several other processes, such as epidemic spreading which are best captured by networks exhibiting multiplex structure \cite{epimul1,epimul2}. Among several dynamical states, the existence of explosive synchronization \cite{expl_syn1}, breathing synchronization \cite{breath_syn}, cluster synchrony \cite{cluster_syn1}, inter- and intralayer synchronization \cite{inter_1,inter_2,intra_1,intra_2}, chimera states \cite{mplxchi2,mplxchi1,mplxchi3} in such interdependent networks has been demonstrated earlier.
\par Nevertheless, to the best of our knowledge, all previous studies have contemplated with either completely time-static (or space static) networks with multilayer (multiplex) formalism or networks of spatially moving nodes over a single layer. But, rigid connections between individuals for modeling several interdependent processes, are less practical. Rather mechanisms following which interactions between particulars evolve over time relying on the physical proximity among the individuals, must be employed. In fact, the extremely crucial dynamical consequences arising in a multiplex network of mobile nodes, is yet to be given its due attention. For instance, public transport networks serve as a paradigmatic example of time-varying (spatial) multilayer networks \cite{ptn}. Multilayer (multiplex) also presents a natural framework for analyzing ecological systems \cite{eco} as it particularly allows one to explore influences due to interlayer and intralayer interactions, where mobility can play decisive role in order to bring self-organization \cite{selforg,spat}. Spreading processes through multiple routes of transmission can also be explored best in spatial multiplex formalism \cite{spre1,spre2}. 

\par So, the present article deals with a quite realistic formalism of networks while following a two-layer multiplex formation in which nodes in each layer are allowed to move performing a two-dimensional (2D) lattice random walk and consequently interactions among the entities change over time due to that spatial movement. Owing to this network framework, there are two types of interaction strengths, the interlayer strength $\epsilon$ and intralayer strengths $k_1$ and $k_2$, together with the mobility parameters, such as the vision ranges $\phi_1, \phi_2$ and step sizes $u_1, u_2$ (a detailed description of them are given later). With variations of these network parameters, we witness inter- and intralayer synchronization states in coupled R\"{o}ssler dynamical systems. We present an analytical study on the necessary conditions for the emergence and stability of such synchronization states that excellently match the numerically obtained results. We thoroughly analyze the persistence of interlayer synchrony against both topological (like link based attack in the form of progressive de-multiplexing and also in response to layer node based attacks in terms of successively made static nodes)and dynamical (in terms of single-node basin stability) perturbations. The effects of intralayer mobility parameters in enhancing robustness of interlayer synchronization under both topological and dynamical disturbances are explained. In this context, we would like to emphasize that in most of the earlier works on synchronization, completely percolated single-component networks are mainly dealt with. But the intermittent connectivity pattern in our work does not necessarily always lead to a single-component network. So as far as the impacts of the mobility parameters are concerned, we relate the incarnation of synchronous states to the evolution of giant connected component in the network \cite{eomgcc}.

\par The present work is organized as follows: Sec. II is devoted to the description of the movement algorithm of the layer nodes, whereas Sec. III deals with the formulation of the proposed dynamical network model. The appearance of inter- and intralayer synchrony as a result of variation in different network parameters is discussed in Sec. IV with subsections A, B and C respectively devoted to numerical results, linear stability analysis and discussions on effects of vision range and step size. In Sec. VA and B, we respectively illustrate the survivability of interlayer synchronous state against so-called structural attacks. Dynamical perturbations to the synchronization manifold are explained in Sec. VI. We further discuss the possibility in enhancing the robustness of the synchronized state via variation in the mobility parameters in both Sec. V and VI. Finally in Sec. VII, we provide concluding remarks on the obtained results.

\section{ Mobility in the layer nodes}
This section deals with the explanation of the mechanism through which the nodes in two layers move in a finite region of the two-dimensional (2D) space. To start with, in each of the layers, we randomly place $N$ nodes in a $P\times Q$ node mesh, while associating a position coordinate $(\xi^k_i,\eta^k_i)$ for the $i$-th node of $k$-th layer, $i=1, 2, \cdots , N$ and $k=1,2$. The movement scheme, we undertake, basically follows a 2D-lattice random walk algorithm (which is a generalization of the well-known 1D-lattice random walk on the integer line \cite{latt}) generating a random geometric graph \cite{rgg1,rgg2}, one of the most important and simple models of spatial networks \cite{spat}. A simple change in the position coordinates $(\xi^k_i,\eta^k_i)$  for the $i$-th node after each (and every) time iteration in order to make the nodes to move either to right or left and to up or down, is applied in the following way:

\begin{enumerate}
	\item[a)]  The movements along the positive $x$-axis (the right) or positive $y$-axis (the up) or along the negative $x$-axis (the left) or negative $y$-axis (the down) directions are defined as :
	\begin{equation}
	\begin{array}{lcl}
	\xi^k_i(\tilde t+\delta \tilde t)=\xi^k_i(\tilde t)+u_k~\mbox{cos}\theta^k_i,\\
	\eta^k_i(\tilde t+\delta \tilde t)=\eta^k_i(\tilde t)+u_k~\mbox{sin}\theta^k_i,
	\end{array}
	\end{equation}
	corresponding to the randomly chosen $\theta^k_i \in\{0, \frac{\pi}{2}, \pi, \frac{3\pi}{2}\}$. Here $\delta \tilde t$ is the characteristic time scale for the movement of the nodes.

	 \item[b)] For the next time iteration with a new set of random numbers $\theta^k_i$ for $i=1,2,...,N$ and $k=1,2$, the above adjustments will be applied again for all the nodes etc.
	 	\item[c)] We employ periodic boundary conditions during the procedure so as to make sure that the nodes continue to be in the physical space for all the course of time without being departed.
	 	\end{enumerate}
	 	\par As hinted above, nodes will interact based on the concept of generation of random geometric graphs and hence they will communicate with only those that belong to a specified region (subregion) created for that particular node. For this, inside the physical space in which the nodes are moving, smaller square shaped regions (calling them {\it vision size}) are assigned to every node. Here, we are considering square shaped vision sizes in the direction of the motion of the nodes which is simple to understand and fits well with our {\it on-grid} platform of the node movement in case of a 2D-{\it lattice} random walk. It could have been chosen in other shapes like that of circular or cone shaped vision sizes that has been dealt with in \cite{cone1,cone2,cone3}. But such vision sizes would not allow one to maintain a {\it lattice} random walk.
	 	\par
\begin{figure}[ht]
	\centerline{
		\includegraphics[scale=0.40]{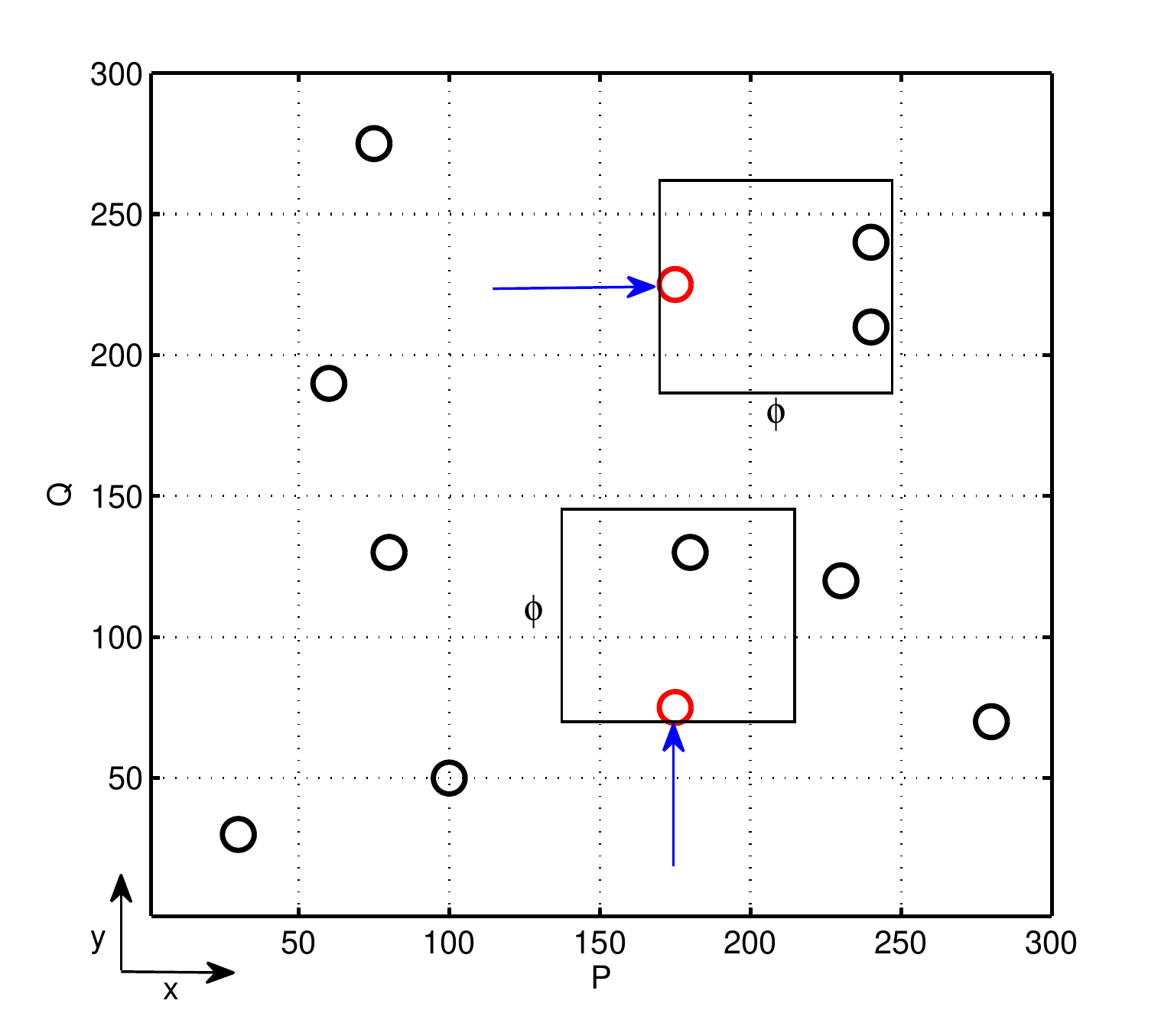}}
	\caption{ Movement in the nodes (denoted by circles) in a $P\times Q$ two-dimensional grid where $P=Q=300$. For clarity of the picture, only $N=12$ moving nodes are shown in a single layer with a view of the vision sizes (having area $\phi^2$) for particularly two nodes (red circles) moving towards positive $x$ and positive $y$ directions (blue arrows).}
	\label{fig10}
\end{figure}

 So, whenever a node moves to its right (say) (cf. Eq. (1)) then a square shaped vision size of area $\phi^2$ is created to its right. Then the node will interact only with those nodes which lie in the vision size. 	 	
	 	Similarly, if a node is moving towards front (cf. Eq. (2)), then a square shaped vision size of area $\phi^2$ is created in that direction. As a matter of fact, the vision size for a particular node will be created in a direction along which the node moves. For a better understanding, a simple graphical view of moving nodes in a single layer and creation of vision sizes at a particular instant of time is presented in Fig. \ref{fig10}.
	 	
	\par Throughout the article, our main emphasis will be to unravel the prominence in the results obtained through the incorporation of mobility in the nodes of the network. For the sake of simplicity, we have considered $\phi_1=\phi_2=\phi$ and $u_1=u_2=u$. Hence we will be mainly concentrating on the impacts of the parameters $\phi$ and $u$ involved in the movement of the nodes. But the frequency of the network switching (node movement) has been considered as the maximum possible here, as at each integration time step the nodes move and the network connectivity gets reshaped. However, the following results have been well verified with lower frequencies in the movement (not shown here) that produce qualitatively similar results.

\section{ Multiplex network of mobile oscillators}
We consider a bi-layer multiplex network with $N$ mobile nodes in each layer performing a 2D-lattice random walk. Each node $i~ (i=1,2,...,N)$ with position coordinate $(\xi_i, \eta_i)$ (cf. Eqs. (1) and (2)) is carrying a dynamical system whose evolution is described by $\dot{X_i}=F(X_i)$, where $X_i$ is a $m$-dimensional vector of dynamical variables of the $i$-th oscillator and $F(X_i)$ is a vector field characterizing the dynamical units. The dynamical network is succinctly given by the following set of equations:
\begin{equation}
\begin{array}{lcl} \label{eq10}
\dot X_{1,i}=F(X_{1,i})-k_1\sum\limits_{j=1}^N g^1_{ij}(t)E_1 (X_{1,j})\\~~~~~~~~~~~~~~~~~~~~~+\epsilon[H (X_{2,i})-H (X_{1,i})],\\
\dot X_{2,i}=F(X_{2,i})-k_2\sum\limits_{j=1}^N g^2_{ij}(t)E_2 (X_{2,j})\\~~~~~~~~~~~~~~~~~~~~~+\epsilon[H (X_{1,i})-H (X_{2,i})],
\end{array}
\end{equation}
where $k_1$, $k_2$ are the intralayer interaction strengths  among the random walkers in layer-$1$ and layer-$2$ respectively, while $\epsilon$ stands for the interlayer coupling strength. Here $G_1(t)=[g^1_{ij}(t)]_{N\times N}$ and $G_2(t)=[g^2_{ij}(t)]_{N\times N}$ are the time-varying zero-row sum Laplacian matrices of order $N$ governing the connectivities in the layers at time $t$.  Particularly, $g^k_{ij}(t)=-1$ if the $j$-th oscillator lies in the vision size of the $i$-th oscillator and otherwise zero, with $k=1,2$. $E_k:\mathbb{R}^m\rightarrow\mathbb{R}^m$and $H:\mathbb{R}^m\rightarrow\mathbb{R}^m$ respectively correspond to the intralayer and interlayer output vectorial functions, $k=1, 2$.

\section{ RESULTS}

Here we are concerned with a bi-layer network where the layer nodes follow the dynamics of well known R\"{o}ssler oscillators \cite{ross_org} with  $F(X_{k,i})$ in the following form: \\
\begin{equation}
F(X_{k,i})=\left(
\begin{array}{c}
-y_{k,i}-z_{k,i}\\
x_{k,i}+ay_{k,i}\\
b+z_{k,i}(x_{k,i}-c)\\
\end{array}
\right), \\
\end{equation}
where $a=0.2, b=0.2, c=5.7$ has been considered so as to keep the oscillators in a chaotic regime. Without loss of generality, we consider both intralayer and interlayer coupling functions to be linear diffusive through the variable $y$, i.e. $E_k(X_k)=(0,y_k,0)'$ and $H(X_k)=(0,y_k,0)'$ where $(\cdot)'$ denotes the transpose of a matrix, with $k=1,2$. Also we fixed the parameters $P=Q=300$, $N=100$ throughout this paper and chose random initial conditions from the phase space volume $[-15, 15]\times [-15, 15]\times [0, 35]$ for the dynamical variables for numerical simulations  \cite{simu}. 

\subsection{Numerical results}

First of all, we define the interlayer and intralayer synchronization errors respectively as follows:
\begin{equation}
E_1=\lim_{T\to\infty} \frac{1}{T}\int_{0}^{T} \sum_{i=1}^{N} \frac{\|\delta W_i(t)\|}{N} dt,
\end{equation}
and
$~~~~~~~~~E_2={\bf e_1}+{\bf e_2}$, with
\begin{equation}
{\bf e_k}=\lim_{T\to\infty} \frac{1}{T}\int_{0}^{T}\sum_{i=2}^{N} \frac{\| X_{k,i}(t)- X_{k,1}(t)\|}{N-1}dt;~~~k=1,2
\end{equation}
where $\delta W_i(t)= X_{1,i}(t)-X_{2,i}(t)$ is the variable describing state difference between the $i$-th replica nodes and $\|\cdot \|$ is the Euclidean norm. Basically, $E_1$ defines a long time \cite{simu2} averaged difference between the replica nodes' dynamics of the two layers varying in the range $[0,~0.35]$, the zero value \cite{e1} of which signifies interlayer synchronization. It corresponds to the situation in which each node in one of the layers is synchronized with its replica in the other layer, irrespective of whether or not it is in synchrony with the other dynamical units of that particular layer. On the other hand, $E_2$ is the sum of the intralayer errors that varies here in the range $[0,~0.70]$, whose zero value actually correspond to the state of intralayer synchronization, whereby both layers are individually synchronized.

\begin{figure}[ht]
	\centerline{
		\includegraphics[scale=0.52]{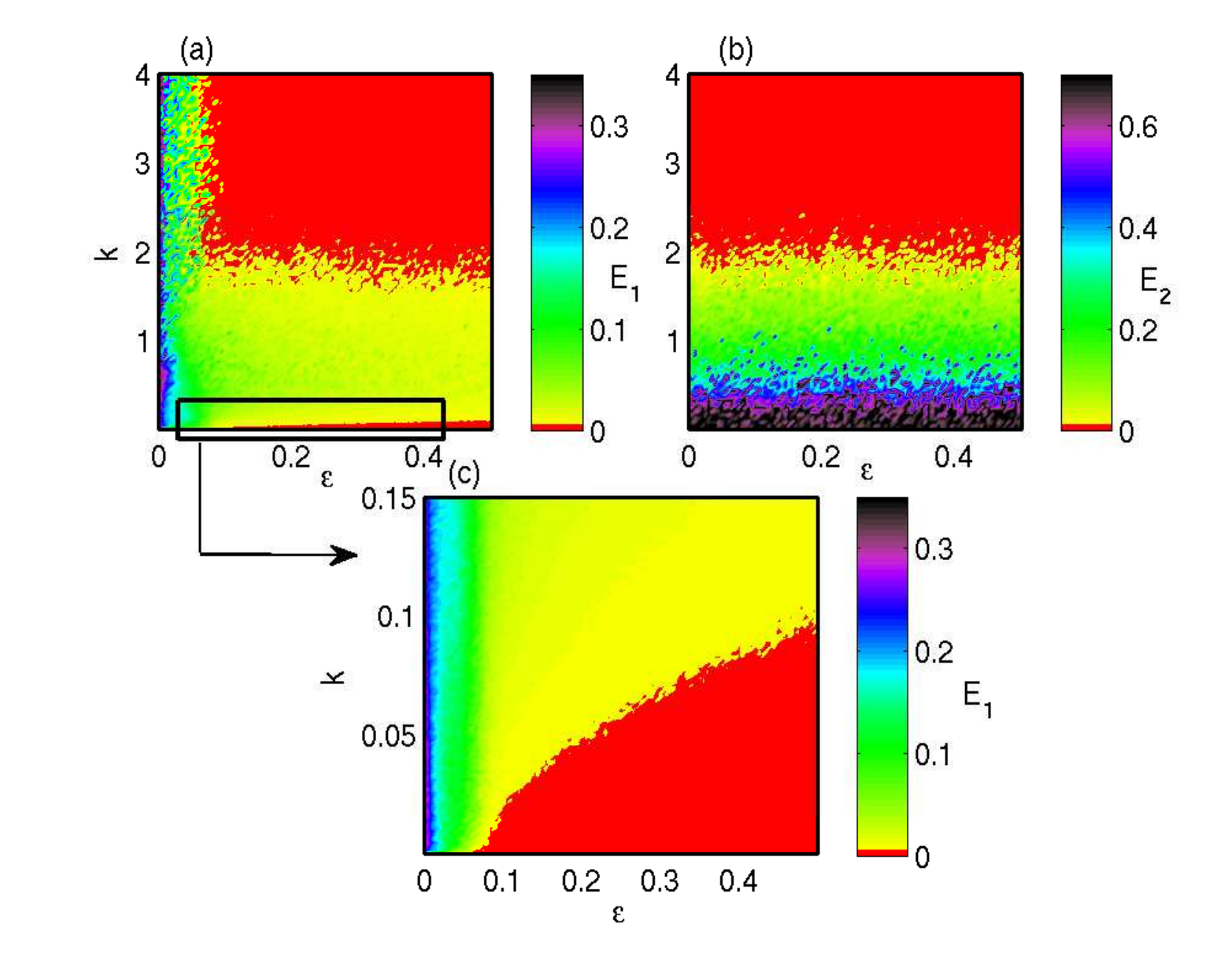}}
	\caption{(a) Interlayer synchronization error $E_1$ and (b) Intralayer synchronization error $E_2$ in the $k-\epsilon$ parameter plane. A magnified version of a specified region in (a) is shown in (c).}
	\label{fig1}
\end{figure}

\par Let us now start by looking at the inter- and intralayer synchronization errors $E_1$ and $E_2$ with respect to simultaneous variation in the interlayer interaction strength $\epsilon$ and strength of intralayer coupling  $k=k_1=k_2$, plotted in Fig.\ref{fig1} obtained through numerical simulations (averaged over 20 realizations) \cite{simu2}. Here we have kept fixed the values of the vision range $\phi=10$ and the step size $u=6$. Figure \ref{fig1}(a) shows that as long as both $k$ and $\epsilon$ are small enough, there is no sign of any kind of synchrony in the network. But, whenever $\epsilon$ crosses the value $\epsilon \simeq 0.10$ with $k$ being small below $k \simeq 0.15$, the network experiences interlayer synchronization satisfying $E_1 \simeq 0$ represented by red color in the figure. However, as $k$ increases, this occurrence of complete interlayer synchrony is getting hindered by intralayer interactions. This observation remains true until $k\simeq 2.0$ above which the perfect interlayer synchrony is witnessed again which persists further. In contrast to this, according to Figure \ref{fig1}(b), intralayer synchrony characterized by $E_2\simeq 0$ appears only when $k\gtrsim 2.0$ and remains for any higher $k$. This readily implies that in the region with small $k\lesssim 0.15$ reflecting interlayer synchrony (discussed above) there is no intralayer synchrony, i.e., synchronization in the replicas may be present even if both layers individually possess asynchrony. But, in the region with $k\gtrsim 2.0$ and $\epsilon\gtrsim 0.10$ both inter- and intralayer synchronizations appear simultaneously. For a better perception of these phenomena, a magnified version of the specified region of Fig.\ref{fig1}(a) is presented in Fig.\ref{fig1}(c) with $0\le \epsilon\le 0.50$ and $0\le k \le 0.15$. From this figure, the values of $E_1$ implying interlayer synchrony are now conspicuous.  Thus, for a fixed $\epsilon$ that exceeds $0.10$, transitions from complete interlayer synchrony to disorder and to perfect interlayer synchrony again can be realized in the network with a variation in the intralayer coupling strength $k$ and this needs further investigations. So, in the next sub-section, we provide analytical results obtained through linear stability analysis to check whether they validate our numerical results.

 \subsection{Linear stability analysis}
\par We proceed through the master stability function (MSF) approach to study the emergence and stability of interlayer synchronization. For this, let $\delta W_i=X_{1,i}- X_{2,i}$ be a small perturbation in the dynamics of the $i$-th nodes of the two layers from the interlayer synchronization manifold $X_{1,i}=X_{2,i}$, for $i=1,2,\cdot \cdot
\cdot , N$. This yield the linearized equation near this manifold (expanding up to the first order) as follows:
\begin{equation}
\begin{array}{lcl} \label{eq15}
\delta \dot W_i=[JF (\tilde{X}_i)-2\epsilon JH (\tilde{X}_i)]\delta W_i-k\sum\limits_{j=1}^{N}g^2_{ij}JE_2( \tilde{X}_j)\delta W_j\\ \hspace{80pt}+ k\sum\limits_{j=1}^{N} \Delta g_{ij} E_2(\tilde{X}_j),
\end{array}
\end{equation}
with $\Delta g_{ij}=g^1_{ij}-g^2_{ij}$ as the difference between the two Laplacians, $J$ stands for the Jacobian and $\tilde{X}_i$ being the state variable corresponding to the isolated $i$-th node following:
\begin{equation}
 \begin{array}{lcl} \label{eq16}
 \dot {\tilde{X}}_i=F(\tilde{X}_i)-k\sum\limits_{j=1}^{N} g^1_{ij}E_1(\tilde{X}_j).
 \end{array}
 \end{equation}
 
  Here the final term on the right-hand side of Eq. \ref{eq15} drives the system away from the interlayer synchronization manifold whenever the interaction topologies of the two layers differ significantly. However, it is expected that when the difference between the two topologies is very small, the MSF approach is applicable \cite{inter_2}. As clarified earlier, the mobility parameters together with the interaction strengths of the individual layers have been taken the same in the present article so that the two layers remain {\it statistically equivalent}. This helps in making the difference $\Delta g_{ij}$ appearing in the last term of Eq.(\ref{eq15}) small so that the negativity of the maximum Lyapunov exponent (MLE) $\Lambda_1$ from Eqs.(\ref{eq15}) and (\ref{eq16}) corresponds to stable interlayer synchrony as the perturbation transverse to the manifold vanishes.
  \par On the other hand, for intralayer synchrony, let the manifolds of the two layers be $S_1(t)=X_{1,i}(t)$ and $S_2(t)=X_{2,i}(t)$, for $i=1,2,...,N$ where $\delta X_{1,i}(t)$ and $\delta X_{2,i}(t)$ are the deviations from the respective manifolds implying $X_{1,i}(t)= S_1(t)+\delta X_{1,i}(t)$ and $X_{2,i}(t)= S_2(t)+\delta X_{2,i}(t)$. Then the perturbed linearized equations become
\begin{equation}
\begin{array}{lcl} \label{eq18}
\delta \dot X_{1,i}=JF(S_1)\delta X_{1,i}-k\sum\limits_{j=1}^{N}g^1_{ij}JE_1(S_1)\delta X_{1,j}\hspace{60pt}\\\hspace{100pt}+\epsilon[JH(S_2)\delta X_{2,i}-JH(S_1)\delta X_{1,i}]
\end{array}
\end{equation}
and
\begin{equation}
\begin{array}{lcl} \label{eq19}
\delta \dot X_{2,i}=JF(S_2)\delta X_{2,i}-k\sum\limits_{j=1}^{N}g^2_{ij}JE_2(S_2)\delta X_{2,j}\hspace{60pt}\\\hspace{100pt}+\epsilon[JH(S_1)\delta X_{1,i}-JH(S_2)\delta X_{2,i}],
\end{array}
\end{equation}
with $(\tilde{X}_{1,i}, \tilde{X}_{2,i})$ as the state variables of the system
\begin{equation}
\begin{array}{lcl} \label{eq20}
\dot{\tilde{X}}_{1,i}=F(\tilde{X}_{1,i})+\epsilon[H(\tilde{X}_{2,i})-H(\tilde{X}_{1,i})]\\
\dot{\tilde{X}}_{2,i}=F(\tilde{X}_{2,i})+\epsilon[H(\tilde{X}_{1,i})-H(\tilde{X}_{2,i})].
\end{array}
\end{equation}

\begin{figure}[ht]
  	\centerline{
  		\includegraphics[scale=0.45]{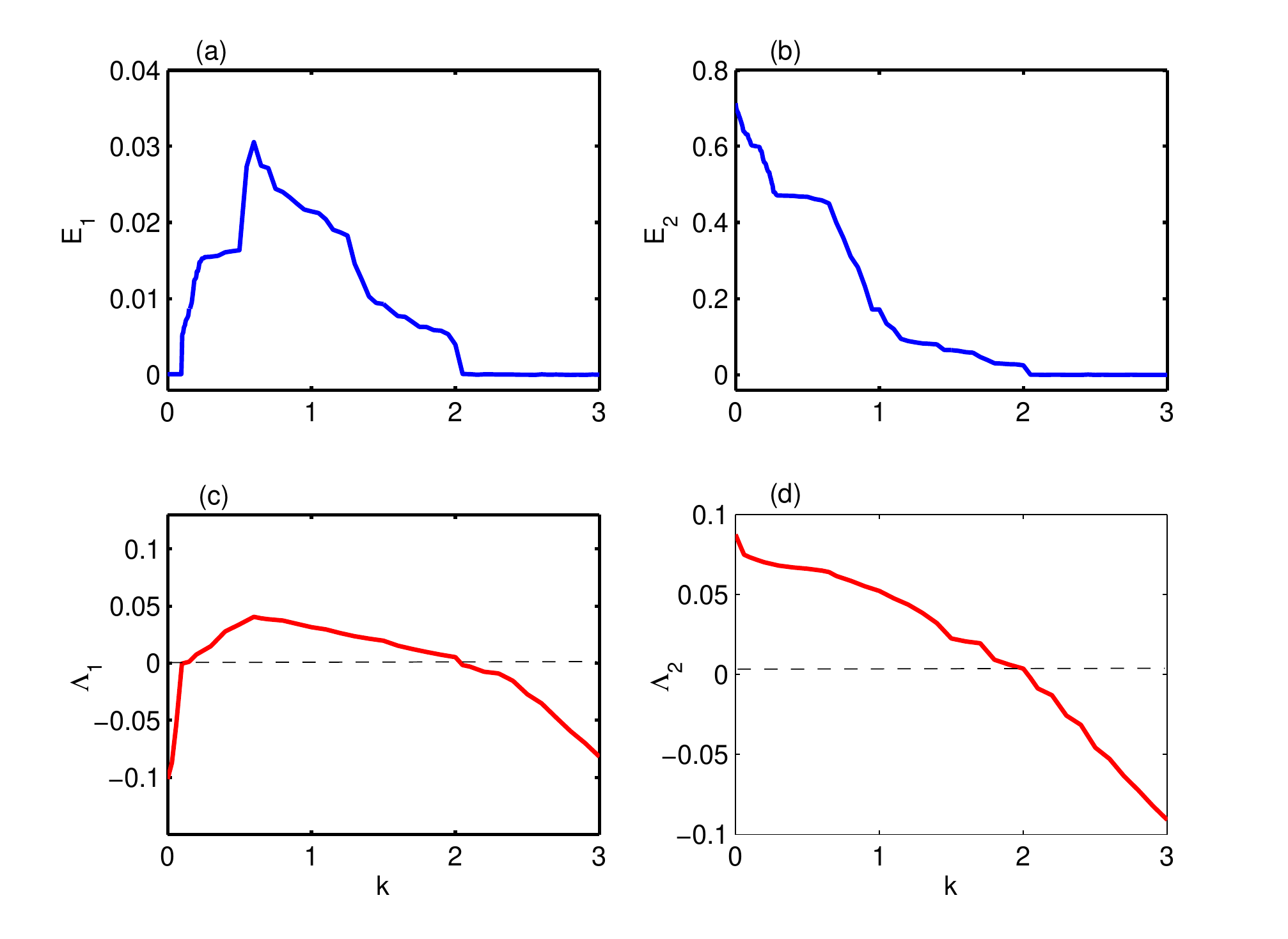}}
  	\caption{ (a,b) Inter and intralayer synchronization errors $E_1$ and $E_2$ respectively; (c,d) Maximum Lyapunov exponents $\Lambda_1$ and $\Lambda_2$ respectively for inter- and intralayer synchronization, all with respect to the parameter $k$ where $\epsilon=0.50$. The other parameters are $\phi=10$ and $u=6$. }
  	\label{fig2}
  \end{figure}

Next solving the linearized systems Eqs.(\ref{eq18}) and (\ref{eq19}) along with the nonlinear Eq.(\ref{eq20}), two maximum Lyapunov exponents transverse to the intralayer synchronization manifolds $S_1(t)=X_{1,i}(t)$ and $S_2(t)=X_{2,i}(t)$ are computed. Then the negativity of the maximum value ($\Lambda_2$) between these two MLE's characterizes the existence and stability of the intralayer synchronization as this situation reflects the occurrence of complete synchronization of the individual layers. In the following, we vary the intralayer coupling strength $k$ (with fixed values of $\phi$ and $u$ which do not appear explicitly in the MSF but they actually govern the Laplacians $g^k_{ij}$ in the equations) and observe the variations in the errors as well as in $\Lambda_1$ and $\Lambda_2$.
\par Figures \ref{fig2}(a) and (b) respectively show the inter- and intralayer synchronization errors $E_1$ and $E_2$ with respect to the intralayer coupling strength $k$, while the interlayer interaction strength is kept fixed at $\epsilon=0.50$. Initially, $E_1$ retains its value to $E_1\simeq 0$ until $k\simeq 0.10$. Then $E_1$ starts varying and remains non-zero in the range $0.10\le k \le 2.05$ signifying the state of out-of synchrony. However for $k>2.05$, $E_1$ becomes zero again implying the re-occurrence of interlayer synchronization. On the contrary, starting with non-zero values, $E_2$ turns into zero only when $k$ passes the value $k=2.05$ and remains zero for higher $k$. This whole scenario is then verified with the values of $\Lambda_1$ and $\Lambda_2$ calculated following the procedure described above. As in Fig.\ref{fig2}(c), $\Lambda_1$ remains negative as long as $k\le 0.10$, but, it starts increasing beyond zero as $k$ increases and $\Lambda_1$ stays positive for $0.10\le k \le 2.05$. Thereafter $\Lambda_1$ is observed to be negative confirming the emergence and stability of interlayer synchrony. Moreover, the change in $\Lambda_2$ is also depicted in Fig.\ref{fig2}(d) where $\Lambda_2$ initializing from positive values, it turns to be negative whenever $k$ crosses the value $k=2.05$ that validates the appearance of intralayer synchronization.

\subsection{Impacts of vision range and step size}

\begin{figure}[ht]
 	\centerline{
 		\includegraphics[scale=0.50]{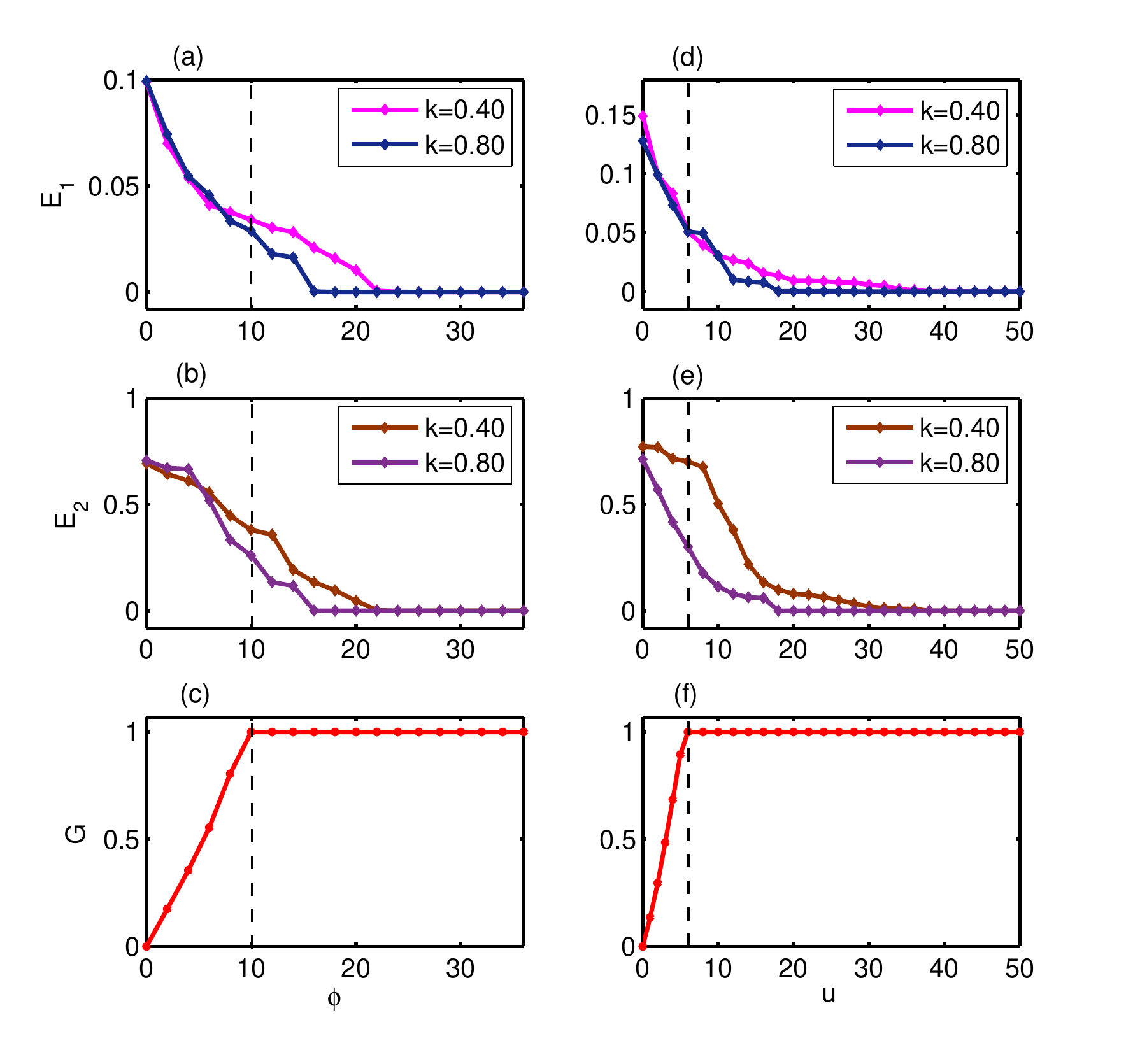}}
 	\caption{(a,d) Interlayer and (b,e) intralayer synchronization errors $E_1$ and $E_2$ together with (c,f) the normalized size $G$ of the giant connected component against $\phi$ (the left panel) with $u=6$ fixed and against $u$ (the right panel) having $\phi=10$ fixed. Here $\epsilon=0.20$. The vertical lines in the two panels correspond to the values $\phi=10$ and $u=6$ which are used in Fig. \ref{fig1}.} 
 	\label{fig3}
 \end{figure}

So far, the results discussed are mainly due to the variation in the two types of interaction strengths $k$ and $\epsilon$ in which the influences of either the vision range $\phi$ or the step size $u$ has not been tested yet. But as we are dealing with a special network formalism of mobile nodes, the consequences of variation in $\phi$ and $u$ deserve attentive study. The errors $E_1$ and $E_2$ are plotted against increasing $\phi$ for two different $k=0.40$ and $k=0.80$, while keeping all the other network parameters fixed at $\epsilon=0.20$ and $u=6$, in Figs.\ref{fig3}(a) and (b). As can be seen in Fig. \ref{fig3}(a) for $k=0.40$, there is no sign of interlayer synchrony (as, $E_1$ is not near zero) till $\phi<22$. But quite interestingly even though $\phi$ actually regulates the intralayer movements, when $\phi$ increases beyond this value the network experiences interlayer synchronization. For higher intralayer coupling strength $k=0.80$, again from non-zero values, $E_1$ turns out to be zero for $\phi \ge 16$. A similar transition has been realized in $E_2$ for increasing $\phi$. With both $k=0.40$ and $k=0.80$, $E_2$ decreases quite rapidly due to the increment in $\phi$ and eventually becomes zero respectively at $\phi=22$ and $\phi=16$ meaning the emergence of intralayer synchrony.
\par Next we identify the normalized (relative) size $G$ (obtained through dividing the actual size by the total size $N$ of the network) of the giant connected component \cite{gc,gc2}, while taking the multiplex network as an ordinary network by averaging over $500$ time units and over $10$ realizations. We plot $G$ as a function of $\phi$ in Fig. \ref{fig3}(c) that explains how $G$ starts increasing rapidly with $\phi$ and eventually aprroaches unity for $\phi\ge 10$, so that the network gets settled into a single component. This percolation is followed by the emergence of synchronizations as depicted in Fig. \ref{fig3}(a,b). Of course, it depends on the value of $k$, the higher the value of $k$, the lower the $\phi$ is needed.

We then want to go for studying the effect of the step size $u$ on these two types of synchronization phenomena. For this, we have chosen $k=0.40$ and $k=0.80$ with $\epsilon=0.20$ as above and kept $\phi=10$ fixed. Figure \ref{fig3}(d) depicts the values of $E_1$ with respect to variation in the step size $u\ge 0$. From this, one is able to see that $E_1$ goes to zero at $u=38$ whenever $k=0.40$ and while $k=0.80$ interlayer synchrony appears for $u=18$. This means changing only the step size of the nodes in the layers, one can achieve interlayer synchronization. Besides, we have also plotted $E_2$ versus $u$ for $k=0.40$ and $0.80$ in Fig.\ref{fig3}(e) where again a similar sort of transition has been observed. The relative size $G$ of the giant connected component with respect to $u$ is plotted in Fig. \ref{fig3}(f) in which increasing $G$ reaches to the unit value for $u\ge 6$ after which synchronization occurs.

\begin{figure}[ht]
	\centerline{
		\includegraphics[scale=0.57]{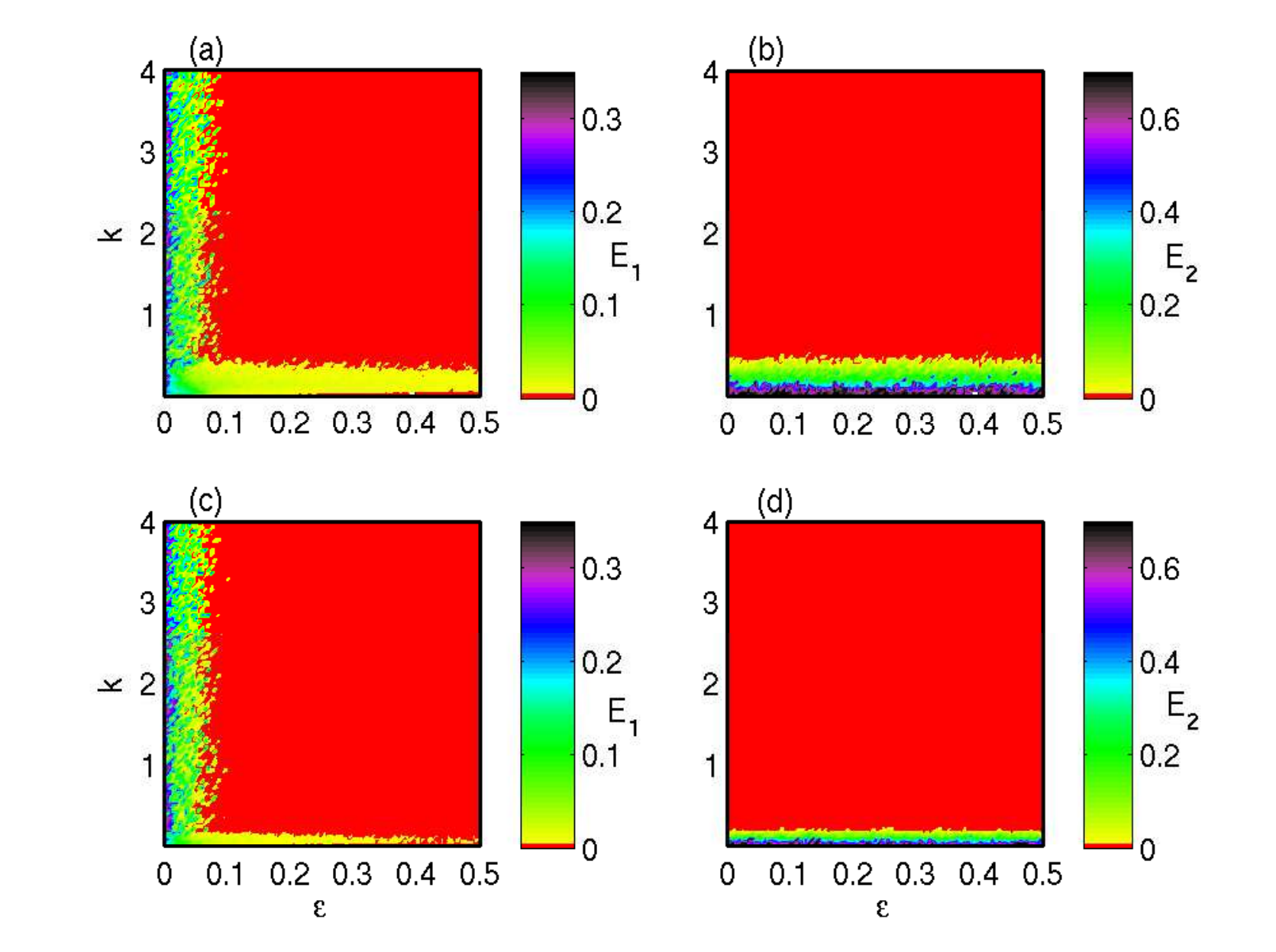}}
	\caption{(a,c) Interlayer synchronization error $E_1$ and (b,d) Intralayer synchronization error $E_2$ for simultaneous variation of $k$ and $\epsilon$. Here $\phi=20,~u=10$ for (a,b) and $\phi=30,~u=15$ for (c,d). }
	\label{fig4}
\end{figure}

\par For an outright visualization of this influence of the mobility parameters $\phi$ and $u$ on the synchronized states, we further went to plot the errors $E_1$ and $E_2$ in the $k-\epsilon$ coupling parameter plane. The errors $E_1$ and $E_2$ are respectively shown in Figs. \ref{fig4}(a) and (b) in the $k-\epsilon$ plane for $\phi=20$ and $u=10$ fixed. Similar scenarios can be realized here as found in Figs.\ref{fig1}(a) and (b). For very small $k\le 0.05$, increasing $\epsilon$ may induce interlayer synchrony, but higher $k\le 0.45$ can disturb this as well. However this is not the case for intralayer synchronization. Above $k\simeq 0.45$ (which was $k\simeq 2.0$ for $\phi=10$ and $u=6$) both inter- and intralayer synchronization take place in the network. With higher $\phi=30$ and $u=15$, both synchronized states appear even earlier beyond $k\simeq 0.15$. This is how the intralayer mobility parameters $\phi$ and $u$ help in not only forming intralayer ordering, but also they are quite effective in creating interlayer synchronized activities.
\par
\begin{figure}[ht]
	\centerline{
		\includegraphics[scale=0.55]{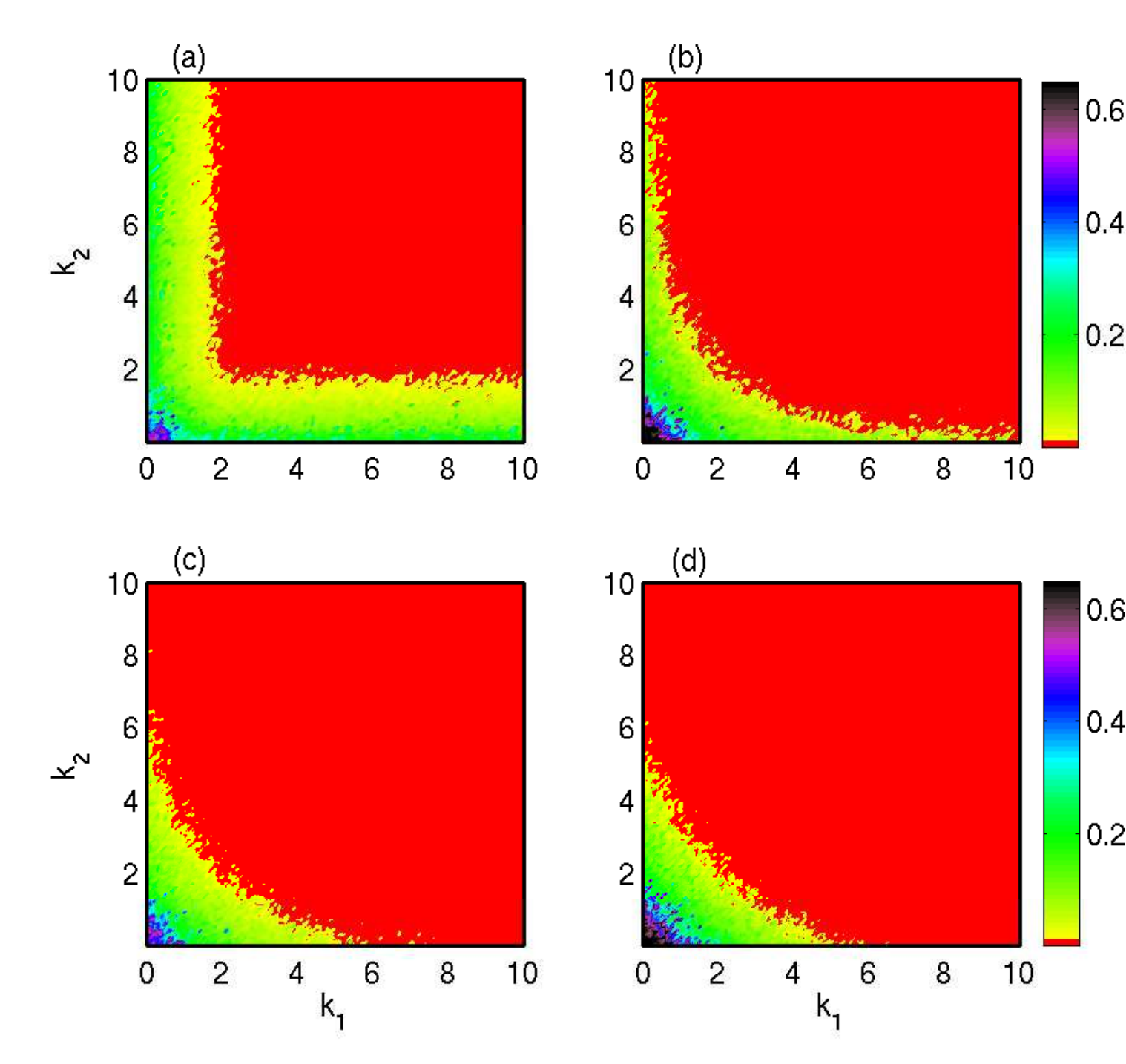}}
	\caption{ Intralayer synchronization error $E_2$ in the $k_1-k_2$ parameter plane with (a) $\epsilon=0.01$, (b) $\epsilon=0.05$, (c) $\epsilon=0.20$, (d) $\epsilon=0.40$. Other parameters are $\phi=10$ and $u=6$. }
	\label{fig5}
\end{figure}

 To have an additional view of the effect of alteration in the interlayer coupling strength $\epsilon$ on the intralayer synchronization, we plot the error $E_2$ in the $k_1-k_2$ plane (cf. Fig. \ref{fig5}). Here the two layers possess two different interaction strengths $k_1$ and $k_2$. We start with a small $\epsilon=0.01$ in Fig. \ref{fig5}(a) where intralayer synchrony does not show up unless the coupling strengths satisfy $k_1\gtrsim 2.0$ and $k_2\gtrsim 2.0$. But with a higher $\epsilon=0.05$, the region reflecting intralayer synchrony gets enlarged in the $k_1-k_2$ plane, as can be seen from Fig. \ref{fig5}(b). With higher $\epsilon=0.20$ and $\epsilon=0.40$, the area of synchrony becomes even larger, as discernible from Figs. \ref{fig5}(c) and (d). This establishes the ability of interlayer interaction strength in enhancing intralayer synchronization as well.

\section{Topological perturbation}
\subsection{Influence of de-multiplexing}

For further understanding about the development of synchronization on such an important framework of the network, now we will be dealing with the issue of robustness of synchronization, particularly interlayer synchrony (as this is one of the most non-trivial behavior arising under the multiplex formalism) against progressive random de-multiplexing (as a form of perturbation in the links). The continuous de-multiplexing is done through sequentially removing (randomly chosen) interlayer links one by one until the two layers become entirely isolated \cite{inter_1}.

\begin{figure}[ht]
	\centerline{
		\includegraphics[scale=0.55]{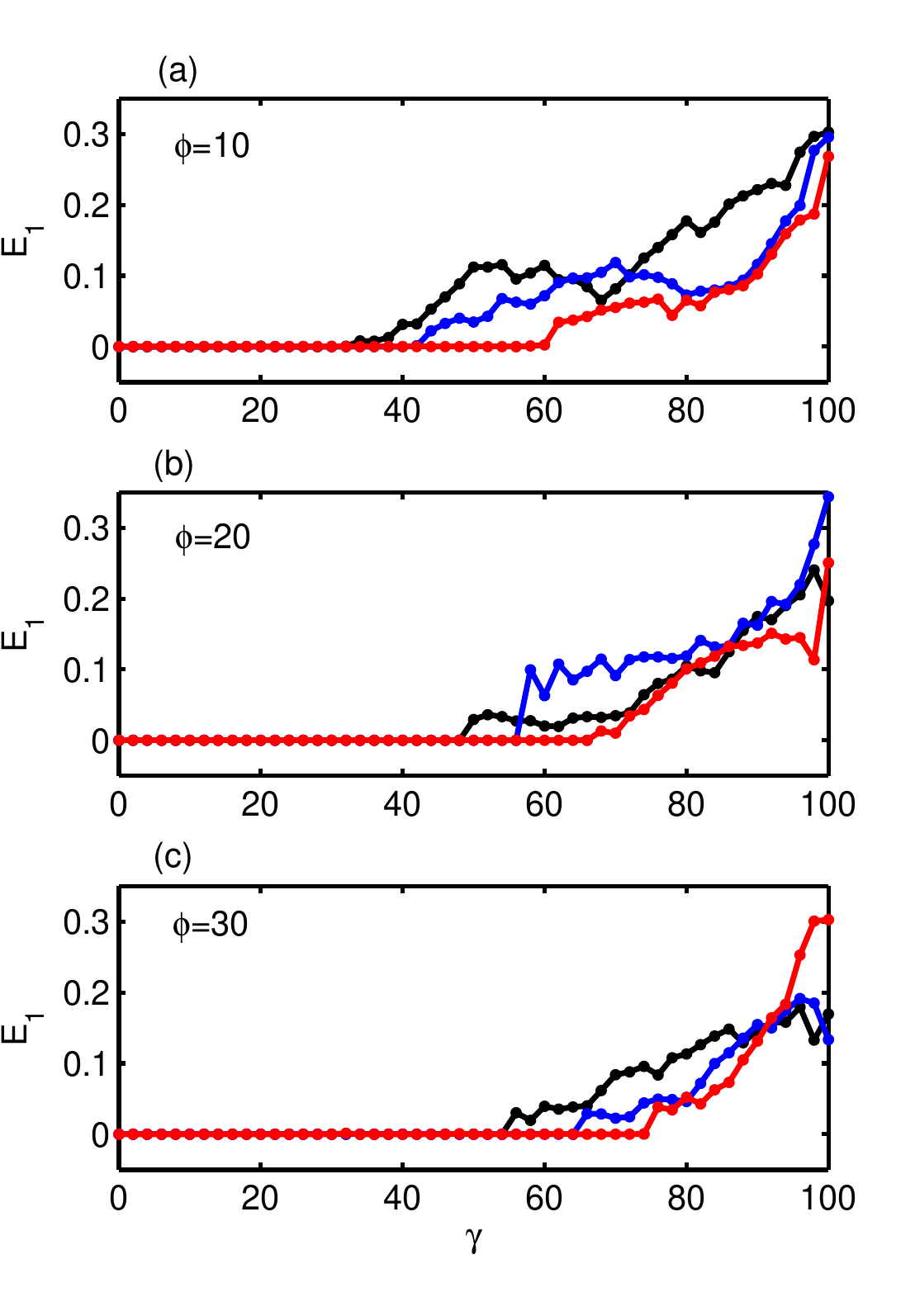}}
	\caption{ Interlayer synchronization error $E_1$ as function of the number of de-multiplexed nodes $\gamma$ for three different values of $k=3.0,~10.0$ and $20.0$ in black, blue and red respectively, with (a) $\phi=10$, (b) $\phi=20$ and (c) $\phi=30$. Here $u=6$. }
	\label{fig6}
\end{figure}

Figure \ref{fig6}(a) shows the interlayer synchronization error $E_1$ against the number of de-multiplexed nodes $\gamma$ starting with $0$ to $N(=100)$. Here, fixed interlayer coupling strength $\epsilon=0.20$ and different values of the intralayer interaction strength $k$ are dealt with, for $\phi=10$ and $u=6$. As we begin with $k=3.0$, one can see that $E_1$ starting from zero, suddenly becomes non-zero whenever $\gamma=32$. This essentially means that the interlayer synchrony is unable to survive for de-multiplexing of $\gamma=32$ replica nodes, but continues to be present even when $31$ pairs of nodes are de-multiplexed. This is because as long as there are possibilities of (indirect) communication between the de-multiplexed pairs of nodes, interlayer synchrony will be there depending on the strength of coupling the nodes are tied with. This can be further illustrated by choosing higher values of $k$. Interestingly, as the intralayer coupling strength $k$ is taken larger than that in the previous case, namely with $k=10.0$, the synchrony persists until $\gamma=42$ (the curve in blue). This shows what a crucial role the intralayer connectivity in terms of $k$, is playing here. Even higher $k=20.0$ makes this critical value of $\gamma$ as $\gamma=57$ (the curve in red) that maintains interlayer synchrony keeping $E_1=0$.

\par Next we investigate the effect of one of the mobility parameter, namely the vision range $\phi$ on this de-multiplexing and consequently the persistence of the interlayer synchrony. In Fig. \ref{fig6}(b), we plot $E_1$ with respect to $\gamma$ for $\phi=20$, while keeping all the other parameters fixed as above. Remarkably, with this $\phi$ and $k=3.0$, $E_1$ remains zero until the number of de-multiplexed nodes exceeds $\gamma=48$ which was previously $\gamma=32$ when $\phi=10$ was taken. Thus for larger $\phi$, interlayer synchrony persists even when the interlayer interaction has been perturbed between a large number of replica nodes. Whenever $k=10.0$ and $k=20.0$ is considered, synchrony survives up to $\gamma=56$ and $\gamma=66$ respectively which were $\gamma=42$ and $\gamma=57$ earlier.
These figures demonstrate that as long as the nodes in each layer get the opportunity of more interactions with other nodes through the generation of larger vision ranges by increasing the value of $\phi$, they can rescue the interlayer synchrony of the network. This scenario is further validated taking $\phi=30$ into account in Fig. \ref{fig6}(c). Here, the critical values of $\gamma$ that retain synchrony, increase over and above and become $\gamma=54, 64$ and $74$ for $k=3.0, 10.0$ and $20.0$ respectively.

\begin{figure}[ht]
	\centerline{
		\includegraphics[scale=0.40]{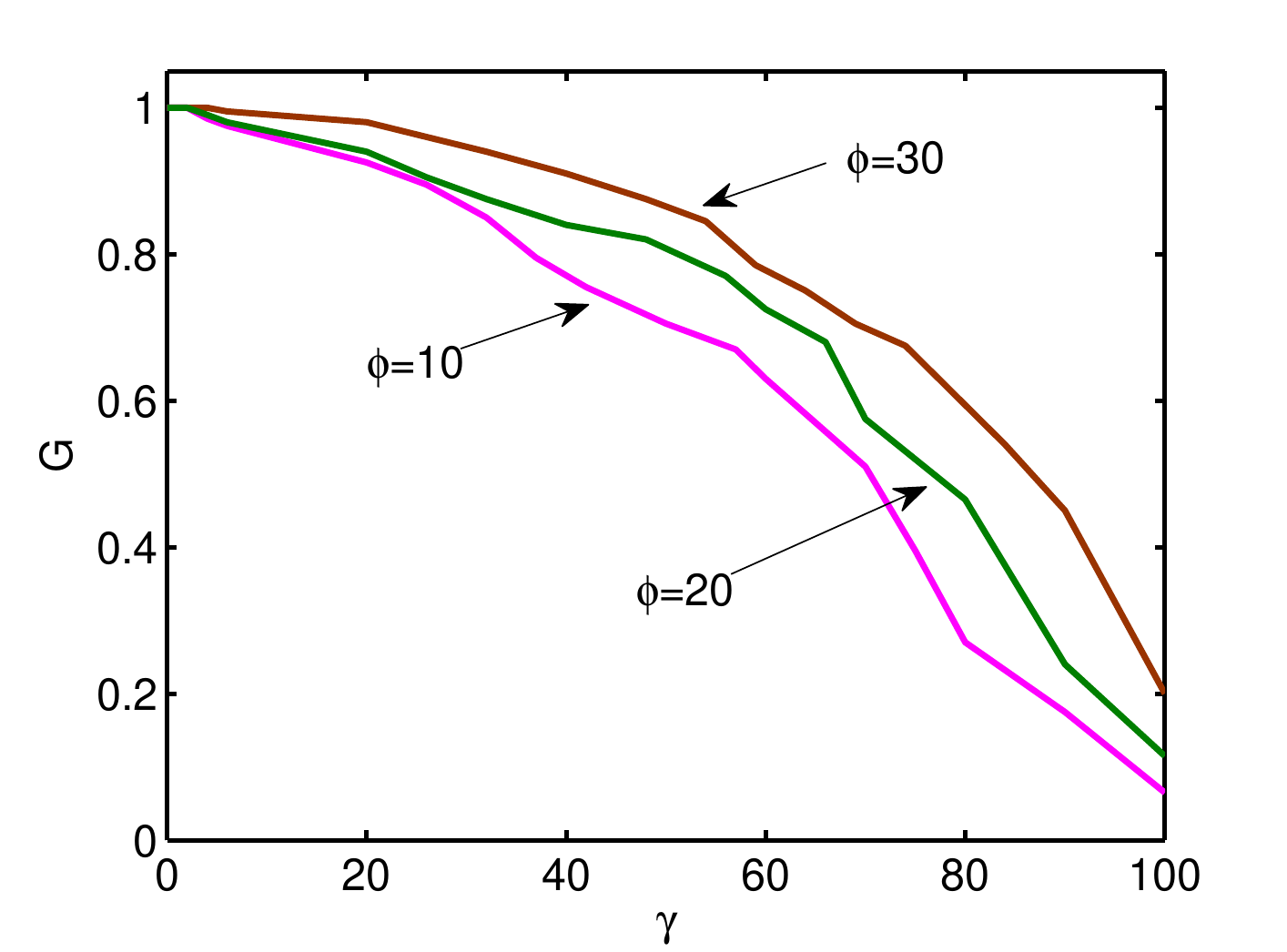}}
	\caption{Normalized size $G$ of the giant connected component with respect to the number of de-multiplexed nodes $\gamma$ for three different values of the vision range $\phi=10$, $\phi=20$ and $\phi=30$. }
	\label{fig7b}
\end{figure}

\par Figure \ref{fig7b} depicts the evolution of the giant connected component against the progressive de-multiplexing in the network and $G$ is plotted with respect to $\gamma$ for the same set of values of $\phi \in \{10,~20,~30\}$ as in Fig.\ref{fig6}. Whenever $\phi=10$, starting with unit value for $\gamma=0$, it decreases monotonically with increasing $\gamma$ depending on which after certain values (of course, it depends on the coupling strength) of $\gamma$ \cite{gc3}, the network loses interlayer synchrony. A similar trend in $G$ for other $\phi$'s is observed but more importantly as we deal with higher vision ranges $\phi=20$ and $\phi=30$, the values of $G$ are essentially higher (or equal) throughout the whole range of $\gamma \in [0,~N]$, respectively varying in the ranges $[0.12,~1.0]$ and $[0.2,~1.0]$ that helps in sustaining synchrony even when a large number of node pairs are de-multiplexed (cf. Fig. \ref{fig6} (b,c)). 
These results indicate that the interlayer synchronization can survive against significant de-multiplexing if it is compensated by a larger vision range of the nodes that governs intralayer mobility.

\subsection{Effect of static nodes}

\par Next we move on to analyze the survivability of interlayer synchrony against successively created static replica nodes (randomly chosen a pair of replica nodes and prescribed $u=0$) in each layer (as a form of disturbance in the nodes). This study is highly reasonable in many ways as some nodes (e.g. in ecological, social networks) may, indeed, undergo though immobility quite naturally or as a result of environmental influence. We will also examine whether mobility parameters (like $\phi$) have any decisive influence in enhancing the persistence of interlayer synchronization.

\begin{figure}[ht]
	\centerline{
		\includegraphics[scale=0.55]{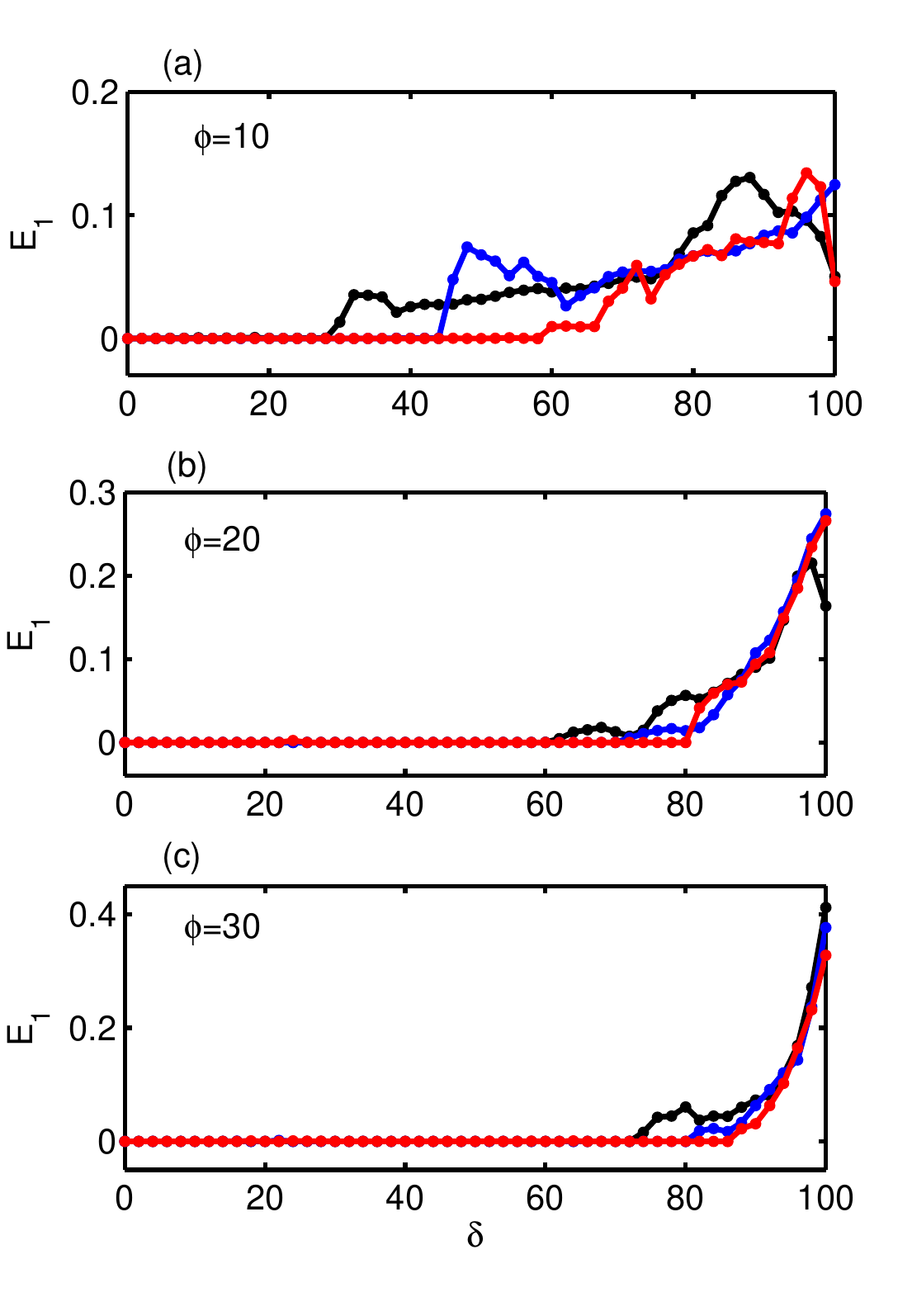}}
	\caption{Interlayer synchronization error $E_1$ with respect to the number of static nodes $\delta$ with three different values of $k=3.0,~6.0$ and $10.0$ in black, blue and red respectively, for (a) $\phi=10$, (b) $\phi=20$ and (c) $\phi=30$. Here $u=6$ is fixed. }
	\label{fig7}
\end{figure}

In Fig. \ref{fig7}(a), $E_1$ is plotted with respect to the number of static nodes $\delta \in [0,N]$ ($\delta=0$ interpreting all nodes are moving, whereas $\delta=N$ means that all the nodes are static in the 2D space for the entire course of time) where $\epsilon=0.20$, $\phi=10$ and $u=6$ are kept fixed. Whenever $k=3.0$, $E_1$ turns non-zero if $\delta=28$ nodes in each layer become static (i.e. $u=0$ for those particular nodes). Until $\delta=28$, $E_1$ remains to be zero. This essentially means that the movement in the rest of the $72$ nodes in each layer is sufficient in order to maintain interlayer synchronization in the network.  Higher $k=6.0$ and $k=10.0$ helps the network to sustain interlayer synchrony even up to $\delta=44$ and $\delta=58$ respectively. So by increasing the intralayer interaction strength $k$, persistence of interlayer synchronization against static replica nodes can be enhanced.   
\par As far as the influence of $\phi$ on this sort of transition is concerned, we assume the vision range $\phi$ to be $\phi=20$ in Fig. \ref{fig7}(b). As seen, this $\phi$ has an amazing impact on the resilience of the interlayer synchrony, as the synchrony can still be present even when $\delta=60$ nodes lose the ability to move any more, for $k=3.0$. With higher $k=6.0$ and $10.0$, this $\phi$ helps the network in experiencing interlayer synchrony until $\delta=70$ and $80$ respectively. Figure \ref{fig7}(c) depicts the enhancement in the sustainability of synchrony in the network with $\phi=30.0$. In this case, $E_1$ remains zero implying the persistence of interlayer synchrony up to $\delta=72, 80$ and $86$ for $k=3.0, 6.0$ and $10.0$ respectively. 
\par
\begin{figure}[ht]
	\centerline{
		\includegraphics[scale=0.40]{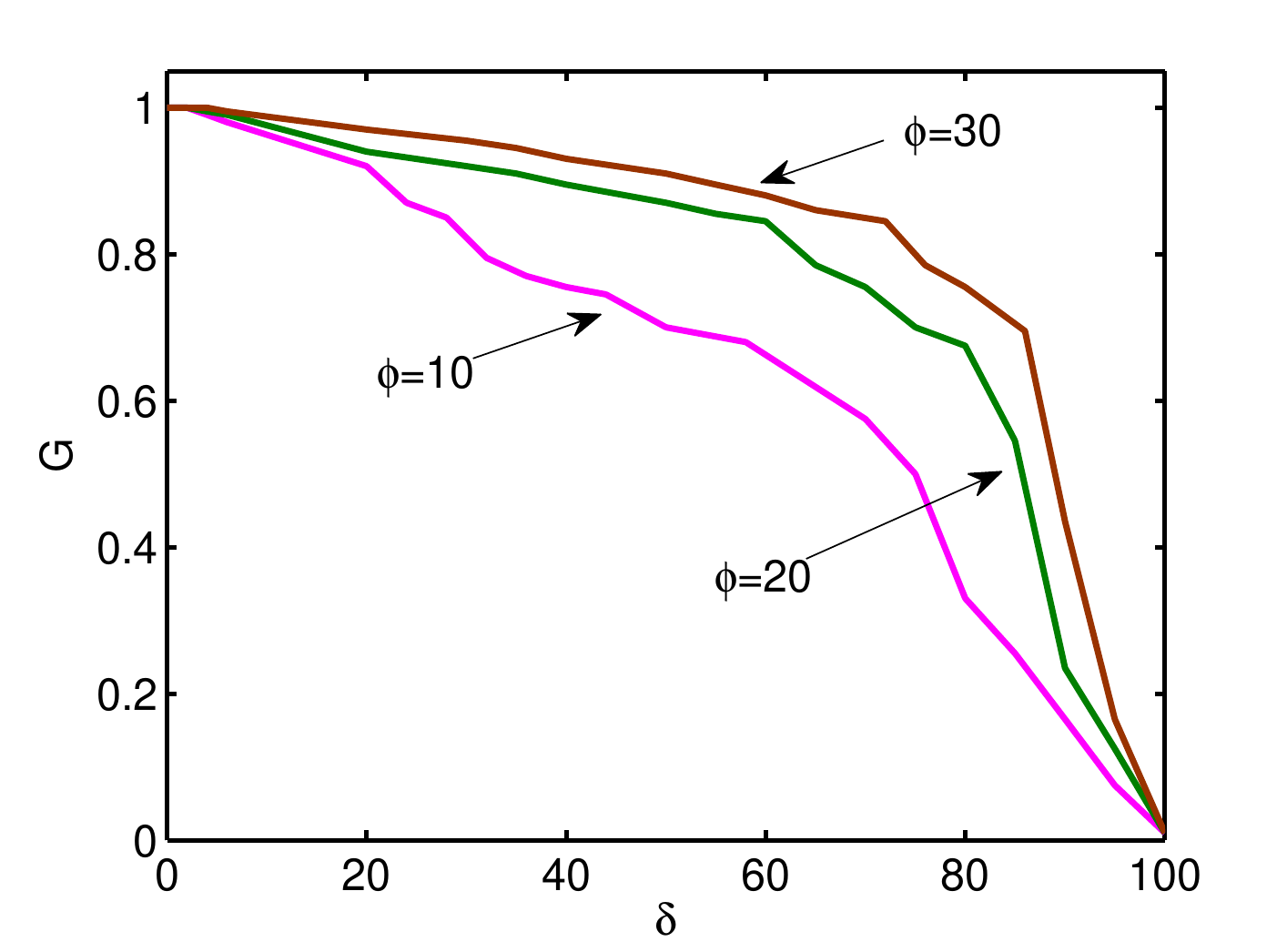}}
	\caption{Normalized size $G$ of the giant connected component as function of the number of static nodes $\delta$ for three different values of the vision range $\phi=10$, $\phi=20$ and $\phi=30$.  }
	\label{fig8b}
\end{figure}

 The variations in $G$ as a function of $\delta$ are figured out in Fig. \ref{fig8b} for these three values of $\phi$. For $\phi=10$, similar to the case of de-multiplexing, starting with unity for $\delta=0$, $G$ decreases for increasing $\delta$ and finally reaches to $G=0.01$ for $\delta=100$, which is the least possible value of $G$ attained when all the nodes become static having no intralayer connections and we are left with only one-to-one interlayer interactions. With $\phi=20$ and $\phi=30$, the nodes get opportunities to have more neighbours which leads $G$ to own higher values (in the same range $[0.01,~1.0]$) and the network to reflect higher extent of robustness. This signifies that if the vision range is high enough then even if a large number of nodes fails to move, the network is still able to realize interlayer synchrony.

\par Such an outcome on augmentation in resilience of synchronization can also be helpful in the study of controlling synchronization which is a very important issue in several fields (e.g. neuronal evolution \cite{neu1,neu2} in which the multiplex formalism \cite{neu3} is evident as well, and some technological \cite{tech1,tech2} and ecological \cite{ec1,ec2} perspectives) and has been a topic of discussion for decades.
\par In this context, we would also like to note that with a different framework of the network (time-static or space-static), it is mainly the interaction strength that governs the evolution of synchronization, but here our study reveals that even if the coupling strength is kept unaltered, the mobility parameters $\phi$ and $u$ will do the rest of the job under consideration.

\section{Dynamical perturbation}
The above two sections were mainly devoted to the study of robustness of interlayer synchrony against some legitimate structural (topological) perturbations and the ability of the network to withstand those failures owing to the disturbance in links and nodes respectively. But the aspect of dynamical perturbation that has high practical relevance, has not been addressed yet. So, in this sub-section, we will go through the discussion of the recently proposed basin stability analysis \cite{bs_nature,bs_delay,bs_nondelay,bs_chimera,mnbs} based on the idea of targeted attacks to specific nodes of the network, termed as ``single-node basin stability (SNBS)" \cite{mnbs}. This scheme of SNBS contributes in understanding the probability of the system to return to its desired stable state after a blow to that particular node through nonlocal arbitrary dynamical perturbations \cite{mnbs2,mnbs3,mnbs4,mnbs5}.
\par The procedure for calculating SNBS is as follows:\\
(i) Obtain the synchronization manifold $S(t)=(X_1,X_2, ..., X_N)$.\\
(ii) Choose $M$ different points that sufficiently move through all parts of the attractor corresponding to the manifold $S(t)$. \\
(iii) Each time initializing the system from a single $m (m=1,2, ..., M)$, one needs to dynamically perturb the $i$-th oscillator by uniformly picking $I$ random initial conditions from the phase space of the networked system.\\
(iv) Among these $I$ initial conditions, count those $J$ initial conditions that reach the synchronized state. Then the SNBS of the $i$-th node starting from the $m$-th point on the attractor is defined as\\
 \begin{equation}
 BS(i,m)=\frac{J}{I},
 \end{equation}\\
(v) Averaging over total $M$-points on the manifold $S(t)$, the mean SNBS can be found as\\
 \begin{equation}
 \langle BS(i)\rangle=\frac{1}{M}\sum\limits_{m=1}^M BS(i,m);~~~~~ (i=1,2, ..., N).
 \end{equation}\\

 \begin{figure}[ht]
 	\centerline{
 		\includegraphics[scale=0.560]{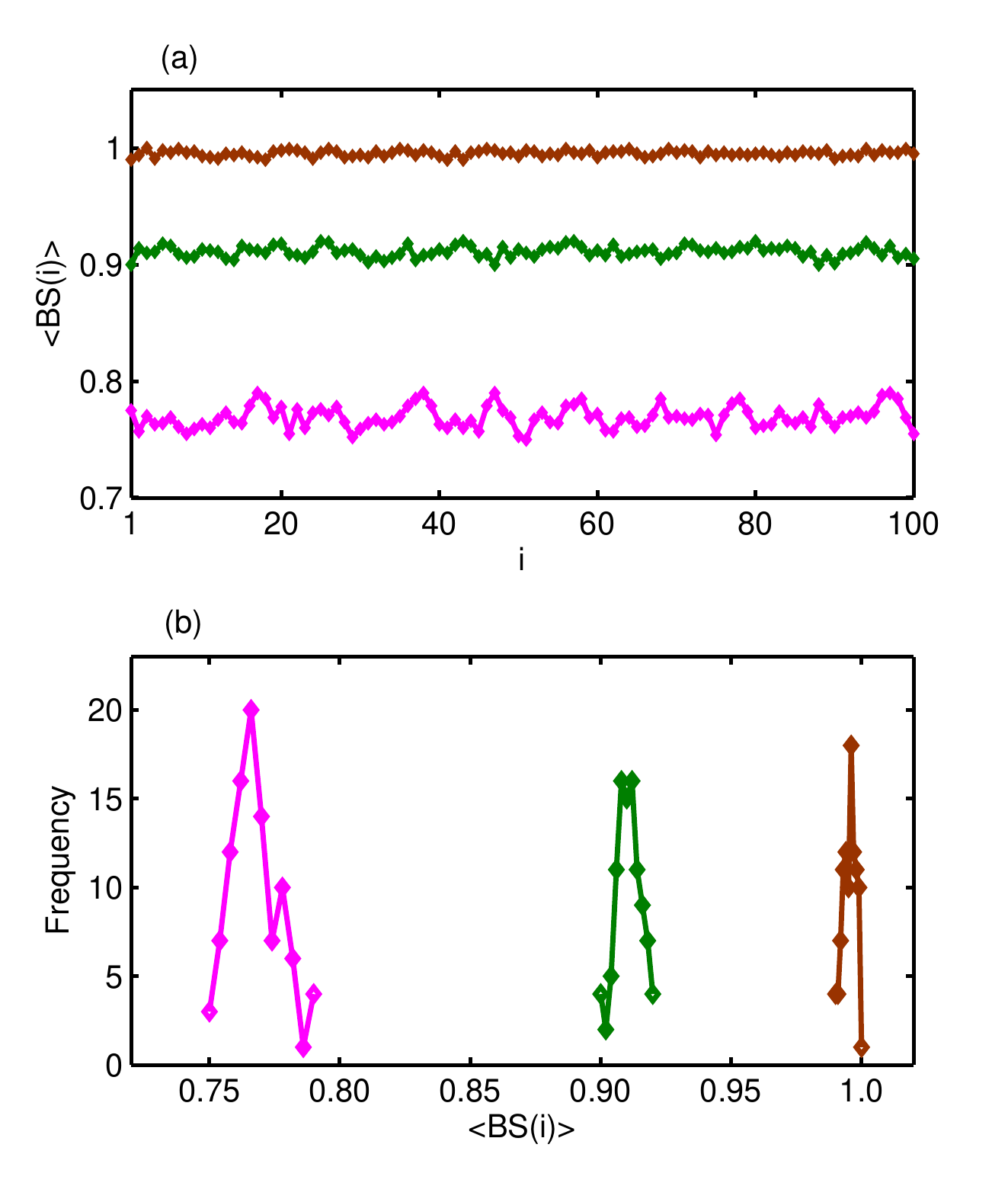}}
 	\caption{Single-node basin stability $ \langle BS(i)\rangle $ of the interlayer synchronization for all the $i$-th pairs of nodes: (a) $\phi=10$ (magenta), $\phi=20$ (deep green) and $\phi=30$ (maroon). The corresponding frequencies of $\langle BS(i) \rangle$ are shown in (b). Other parameters are $k=3.0$, $\epsilon=0.20$ and $u=6$. }
 	\label{fig8}
 \end{figure}

\par Let us now have a look at the estimated values of $ \langle BS(i)\rangle $ of all the pairs of replica nodes $(i=1,2, ..., N)$, where for the computation, we have chosen $I=1000$ random initial conditions from the phase space volume $[-15, 15]\times [-15, 15]\times [0, 35]$ of the individual dynamical units. $M=10$ points on the interlayer synchronization manifold $S(t)$ have been considered.
\par So far, we have observed that whenever $k=3.0, \epsilon=0.20$ with $\phi=10$ and $u=6$, interlayer synchrony has appeared (cf. numerical and linear stability analysis) which is valid as long as the initial conditions are taken sufficiently close to the synchronization manifold. Figure \ref{fig8} shows $ \langle BS(i)\rangle $ of the replica nodes when all the replica nodes are dynamically disturbed from the interlayer synchronization state following the above defined mechanism. As in Fig. \ref{fig8}(a), the values of $ \langle BS(i)\rangle $ for all the $N=100$ pairs of nodes (in magenta) are close to each other remaining in the range $[0.75, 0.79]$. As we noticed in the cases of topological perturbations (cf. Sec. V A and V B), here again due to an increase in the value of the intralayer mobility parameter (namely $\phi$), the network experiences a notable development in the values of $ \langle BS(i)\rangle~~(i=1,2,..., N) $. This time, they lie in the range $[0.90, 0.92]$ (in deep green), as plotted for $\phi=20$. For further raise in $\phi$ to $\phi=30$, almost all the initial states eventually lead to the synchronized state, no matter which pairs of nodes are perturbed, with $ \langle BS(i)\rangle \in [0.99, 1.0]~~(i=1,2,..., N) $ (cf. the curve in maroon). The corresponding frequencies of $ \langle BS(i)\rangle $ for all these three values of $\phi \in \{10,~20,~30\}$, are shown in Fig. \ref{fig8}(b). It is clear that $ \langle BS(i)\rangle $ increases monotonically with $\phi$, while its dispersion has the opposite trend. 

Thus, a sufficient vision range can substantiate an optimal response to perturbations on the nodes. This shows that not only the effects of topological fluctuations but the impacts of dynamical perturbations can also be recompensed through variation in mobility parameters of the network.

\section{Conclusions}
 We have demonstrated here the emergence of one of the most important collective behavior in ensemble of oscillators, i.e., synchronization, in a multiplex dynamical network of mobile nodes. We have considered a two-layer network in which one-to-one correspondence between the replica nodes are preserved, in general. Chaotic dynamics arising from the R\"{o}ssler system is used to cast the nodes of each layer in the network. The nodes in each layer are moving while performing a two-dimensional lattice random walk and regulate their movement after every time iteration. According to the network model presented, there are two coupling parameters, namely the inter- and intralayer interaction strengths $\epsilon$ and $k$ together with the mobility parameters: vision range $\phi$ and the step size $u$. Starting with the discussion of variation in $k$ and $\epsilon$, we have thoroughly investigated the alluring impacts of the intralayer mobility parameters $\phi$ and $u$ in favoring not only the intralayer synchronization but also the interlayer synchrony. Apart from the numerical experiments, necessary conditions for the existence and stability of inter- and intralayer synchrony following master stability function approach has also been provided. For an outright understanding of the survivability of interlayer synchrony, we further applied structural perturbations in the form of progressive de-multiplexing (link-based attack) and sequentially created static nodes (site-based attack). The outcomes suggest that an increment in the intralayer mobility parameter $\phi$ can immensely enhance and resurrect interlayer synchrony in the network from the disordered state.  We have also illustrated the development of the giant connected component to explain these scenarios of synchronization while discussing the issues related to the mobility parameters. Besides this, dynamical perturbation while probing the notion of newly proposed single-node basin stability has also been implemented. Here again such consequence of $\phi$ in reviving interlayer synchrony particularly based on volumes of basin of attraction, is witnessed.
 \par We hope that our findings open up new frontiers in the theory and application of synchronization from various aspects, in particular for situations of individuals' mobility induced complexities.
	\\
	\\
	
\noindent \textbf{Acknowledgments} \\
 D.G. was supported by the Department of Science and Technology, Government of India (Project No. EMR/2016/001039).


\begin{thebibliography}{25}
\bibitem{simu} We integrated the network equation (3) following fifth-order Runge-Kutta-Fehlberg algorithm with a time step size of $h=0.01$.
\bibitem{simu2} For our computation, $T$ is chosen to be $2{\mathrm x}10^5$ time units after an initial transient of $2{\mathrm x}10^5$ time units.
\bibitem{e1} In our simulations, the values of $E_1$ and $E_2$ less than $10^{-5}$ respectively correspond to inter- and intralayer synchronizations.
\bibitem{gc2} This giant connected component is computed for the underlying undirected network, which is usually known as the weakly connected component. The symmetry of the relation of synchronization and the possibility of synchrony through unidirectional coupling sufficiently allows one to calculate the weakly connected component in this context.
\bibitem{gc3} In order to achieve only inter-layer synchronization, the network does not necessarily possess unit value of $G$, smaller values of $G$ may suffice as long as there are possibilities of (indirect) communication between the de-multiplexed pairs of nodes.
\end{thebibliography}
\end{document}